\documentclass[journal]{IEEEtran}

\ifCLASSINFOpdf
\else
\fi

\usepackage{setspace}
\usepackage{bbm}
\usepackage[cmex10]{amsmath}
\usepackage{amssymb}
\usepackage{cite}
\usepackage{graphicx}
\usepackage{array,color}
\usepackage{amsmath}
\usepackage{stfloats}
\usepackage{graphicx}
\usepackage{epstopdf}
\usepackage{subfigure}
\usepackage{tabularx}
\usepackage{epsfig,epsf,color,balance,cite}
\usepackage{algorithmic}
\usepackage{algorithm}
\usepackage{url}
\usepackage{bm}

\usepackage{amsthm}
\ifodd 0
\usepackage{soul}
\usepackage{color}
\setstcolor{red}

\newcommand{\rev}[1]{{\color{red}#1}} 
\newcommand{\del}[1]{\st{#1}} 

\newcommand{\com}[1]{\textbf{\color{red} (COMMENT: #1)}} 
\newcommand{\response}[1]{\textbf{\color{green} (RESPONSE: #1)}} 
\else

\newcommand{\rev}[1]{#1}

\newcommand{\del}[1]{}

\newcommand{\com}[1]{}
\newcommand{\comg}[1]{}
\newcommand{\response}[1]{}
\fi

\title{ {Intelligent Reflecting Surface-Enhanced OFDM: Channel Estimation and Reflection Optimization}}

\author{\normalsize Beixiong Zheng,~\IEEEmembership{Member,~IEEE} and Rui Zhang,~\IEEEmembership{Fellow,~IEEE} 
	\thanks{
		
		The authors are with the Department of Electrical and Computer Engineering, National University of Singapore,
		email: \{elezbe, elezhang\}@nus.edu.sg.

	}
}

\begin{document}
\markboth{IEEE Wireless Communications Letters, Vol. 9, No. 4, April 2020}{SKM: My IEEE article}
\maketitle
\begin{abstract}
In the intelligent reflecting surface (IRS)-enhanced wireless communication system,
channel state information (CSI) is of paramount importance for achieving the passive beamforming gain of IRS, which, however, is a practically challenging task due to its massive number of passive elements without transmitting/receiving capabilities.
In this letter, we propose a practical transmission protocol to execute channel estimation and reflection optimization successively for an IRS-enhanced orthogonal frequency division
multiplexing (OFDM) system.
Under the unit-modulus constraint, a novel reflection pattern at the IRS is designed to aid the channel estimation at the access point (AP) based on the received pilot signals from the user, for which the channel estimation error is derived in closed-form.
With the estimated CSI, the reflection coefficients are then optimized by a low-complexity algorithm based on the resolved strongest signal path in the time domain.
Simulation results corroborate the effectiveness of the proposed channel estimation and reflection optimization methods.
\end{abstract}
\IEEEpeerreviewmaketitle

\vspace{-0.7cm}
\section{Introduction}

\IEEEPARstart{I}{ntelligent} reflecting surface (IRS), which enables the reconfiguration of wireless propagation environment by smartly controlling the signal reflections via its massive low-cost passive elements, has recently emerged as 
a promising new technology for significantly improving the wireless communication coverage, throughput, and energy efficiency \cite{qingqing2019towards,Renzo2019Smart,Huang2018Achievable}.
By jointly adjusting the reflected signal amplitude and/or phase shift at each of the IRS elements according to the dynamic wireless channels, the signals reflected by IRS and propagated through other paths can be constructively combined at the intended receiver to enhance the received signal power. 
Compared to the traditional active relaying/beamforming techniques, 
IRS possesses much lower hardware cost and energy consumption due to passive reflection and yet operates in full-duplex without the need of costly self-interference cancellation \cite{qingqing2019towards}.

However, the enormous passive beamforming gain provided by IRS is achieved at the expense of more overhead for channel estimation in practice, due to the additional channels involved between the IRS and its associated access point (AP)/users. 
Prior works on IRS mainly focus on the design of reflection coefficients under the assumption of perfect channel state information (CSI) \cite{Wu2019TWC}, which facilitates in deriving the system performance upper bound but is difficult to realize in practice. 
In contrast, there has been very limited work on the joint design of practical channel estimation
and reflection optimization under imperfect CSI tailored to the IRS-aided system, especially for wideband communications.
It is worth noting that such design is practically challenging due to the lack of transmitting/receiving as well as signal processing capabilities of the passive IRS elements while their numbers can be practically very large, which thus calls for innovative solutions to tackle these new challenges.

\begin{figure}[!t]
	\centering
	\includegraphics[width=2.5in]{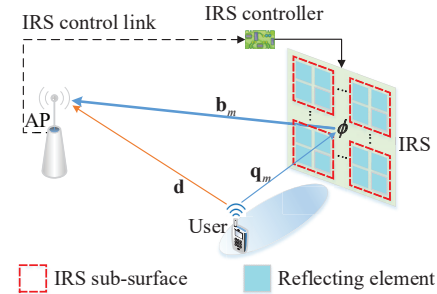}
	\caption{An illustration of IRS-enhanced OFDM communication in the uplink.}
	\label{system}
	\vspace{-0.5cm}
\end{figure}
As compared to the approach of equipping the IRS with dedicated sensors/receiving circuit to enable its channel estimation,
it is more cost-effective to estimate the concatenated user-IRS-AP channels at the AP with properly designed IRS reflection pattern based on the received pilot signals sent by the user and reflected by the IRS \cite{qingqing2019towards}. 
Prior works adopting this method for IRS channel estimation have assumed a simple element-by-element ON/OFF-based reflection pattern \cite{Mishra2019Channel,he2019cascaded,yang2019intelligent}, which, however, has two main drawbacks.
First, it is practically costly to implement the ON/OFF switching of the massive IRS elements frequently as this requires separate  amplitude control (in addition to phase shift) of each IRS element. Second, the large aperture of IRS is not fully utilized as only a small portion of its elements is switched ON at each time, which degrades the channel estimation accuracy.
To overcome the above issues, we propose in this letter a new IRS reflection (phase-shift) pattern for channel estimation by considering the full reflection of the IRS at all time, i.e., all of its elements are switched ON with maximum reflection amplitude during both the channel estimation and data transmission phases. 
As shown in Fig.~\ref{system}, we consider a practical wideband IRS-enhanced orthogonal frequency division
multiplexing (OFDM) system under frequency-selective fading channels, for which a practical transmission protocol is proposed to execute channel estimation and reflection optimization successively.
Specifically, a novel phase-shift pattern satisfying the unit-modulus constraint is designed for the IRS to facilitate the concatenated user-IRS-AP channel estimation at the AP based on the uplink pilot signals from the user.   
A closed-form expression on the channel estimation
error is also derived to show the impact of different system parameters.
Based on the estimated CSI, the reflection coefficients are then optimized to maximize the strongest time-domain path channel gain, which is shown to have a much lower computational complexity as compared to the semidefinite relaxation (SDR) method in \cite{yang2019intelligent} and yet achieve very close performance to it.

\emph{Notation:} 
	Superscripts ${\left(\cdot\right)}^{T}$, ${\left(\cdot\right)}^{H}$, and ${\left(\cdot\right)}^{-1}$ stand for transpose, Hermitian transpose, and matrix inversion operations, respectively.
	$\lfloor \cdot \rfloor$ is the floor function,
	$\odot$ denotes the Hadamard product, and ${\rm rank} \left( \cdot \right)$ denotes the matrix rank,
	and $\angle (\cdot )$ denotes the phase of a complex number.

\vspace{-0.3cm}
\section{System Description and Transmission Protocol}
As illustrated in Fig.~\ref{system},
we consider an uplink OFDM system, where an IRS is deployed to assist in the transmission from a user (in its vicinity) to an AP, both of which are equipped with a single antenna.
Note that the IRS is practically composed of a large number of passive reflecting elements to maximize its reflection power, 
which, however, incurs high overhead/complexity for channel estimation and reflection optimization.
By grouping adjacent elements of the IRS with high channel correlation into a sub-surface to share a common reflection coefficient \cite{yang2019intelligent}, the complexity of channel estimation and reflection design can be significantly reduced.
Accordingly, the IRS composed of $K$ reflecting elements is divided into $M$ sub-surfaces, each of which consists of ${\bar K}=K/M$ adjacent elements, e.g., ${\bar K}=4$ as illustrated in Fig.~\ref{system}.
Moreover, the IRS is connected to a smart controller to enable dynamic adjustment of its elements' individual reflections.
In this letter, quasi-static frequency-selective fading channels are considered for both the user$\rightarrow$AP direct link and the user$\rightarrow$IRS$\rightarrow$AP reflecting link, which remain approximately constant within the transmission frame of our interest.
This is a valid assumption as IRS is practically used to mainly support low-mobility users in its neighborhood only. 
\vspace{-0.3cm}
\subsection{System Model}
With OFDM, the total bandwidth allocated to the user is equally divided into $N$ sub-carriers, which are indexed by $n\in {\cal N} \triangleq \{0,1,\ldots, N-1\}$. For simplicity, we assume that the total transmission power at the user $P_t$ is equally allocated over the $N$ sub-carriers with the power at each sub-carrier given by $p_n=P_t/N,~\forall n \in {\cal N}$.
Without loss of generality, it is assumed that the baseband equivalent channels of both the direct link and the reflecting link have the maximum delay spread
of $L$ taps in the time domain.
At the user side, each OFDM symbol ${\bf x}\triangleq \left[X_0,X_1,\ldots, X_{N-1}   \right]^T$ is first transformed into the time domain via an $N$-point inverse discrete Fourier transform (IDFT), and then appended by a cyclic prefix (CP) of length $L_{cp}$, which is assumed to be  longer than the maximum delay spread of all channels, i.e., $L_{cp}\ge L$. 

At the AP side, after removing the CP and performing the $N$-point discrete Fourier transform (DFT), 
the equivalent baseband received signal in the frequency domain is given by 
\vspace{-0.2cm}
\begin{align}\label{receive}
{\bf y}&={\bf X}\left( \sum_{m=1}^M {\bf q}_{m}\phi_m \odot {\bf b}_{m}+ {\bf d}\right)+
{\bf v}
\end{align}
where 
${\bf y}\triangleq \left[Y_{0}, Y_{1},\ldots,Y_{N-1}   \right]^T$ is the received OFDM symbol,
${\bf X}=\text{diag} \left({\bf x}\right)$ is the diagonal matrix of the OFDM symbol ${\bf x}$,
${\bf d}\triangleq\left[D_{0}, D_{1},\ldots,D_{N-1}   \right]^T\in \mathbb{C}^{ N\times1}$ is the channel frequency response (CFR) of the user$\rightarrow$AP direct link,
${\bf q}_{m}\in \mathbb{C}^{N \times 1}$ is the aggregated CFR of the user$\rightarrow$IRS link associated with the $m$-th sub-surface, 
$\phi_m$ denotes the common reflection coefficient within the $m$-th sub-surface,
${\bf b}_{m}\in \mathbb{C}^{N \times 1}$ is the aggregated CFR of the IRS$\rightarrow$AP link
associated with the $m$-th sub-surface, 
and ${\bf v}\triangleq \left[V_{0},V_{1},\ldots,V_{N-1}   \right]^T \sim {\mathcal N_c }({\bf 0}, \sigma^2{\bf I}_N )$ is the additive white Gaussian noise (AWGN) vector.
In addition, the reflection coefficient $\phi_m$ characterizes 
the equivalent interaction of the $m$-th sub-surface with the incident signal, which can be expressed as \cite{Wu2019TWC}
\vspace{-0.2cm}
\begin{align}
\phi_m= \beta_m e^{j\varphi_m}, \quad m=1, \ldots, M
\end{align}
where $\beta_m \in [0,1]$ and $\varphi_m\in(0, 2\pi]$ stand for the
reflection amplitude and the phase shift of the $m$-th sub-surface, respectively.
To maximize the reflection power of the IRS and simplify its hardware design, we fix $\beta_m=1, \forall m=1, \ldots, M $ and only adjust the phase shift $\varphi_m$ for both channel estimation and reflection optimization in this letter.

By denoting ${\bf g}_{m} \triangleq \left[G_{m,0}, G_{m,1},\ldots, G_{m,N-1}   \right]^T=
{\bf q}_{m}\odot {\bf b}_{m}$ as the equivalent cascaded CFR of the reflecting link without the effect of phase shift for the $m$-th sub-surface, (\ref{receive}) can be rewritten as
\vspace{-0.3cm}
\begin{align}
{\bf y}&={\bf X}\left( \sum_{m=1}^M \phi_m {\bf g}_{m}+ {\bf d}\right)+{\bf v}\label{receive2}
\end{align}
which dispenses with the explicit knowledge of ${\bf q}_{m}$ and ${\bf b}_{m}$ for the reflection design.
Moreover, by stacking $ {\bf g}_{m}$ with $m=1, \ldots, M $ into 
${\bf G}=\left[ {\bf g}_{1}, {\bf g}_{2},\ldots, {\bf g}_{M} \right]$ as the equivalent cascaded CFR matrix of the reflecting link, (\ref{receive2}) can be written in a compact form as
\vspace{-0.3cm}
\begin{align}
{\bf y}={\bf X}\underbrace{ \left( {\bf G} {\bm \phi}+ {\bf d}\right)}_{{{\bf h}}}+{\bf v}\label{receive3}
\end{align}
where ${\bm \phi} \triangleq \left[\phi_1, \phi_2, \ldots, \phi_M \right]^T$ denotes the phase-shift vector and
${{\bf h}}=\left[{H}_{0},{H}_{1},\ldots, {H}_{N-1}   \right]^T$ stands for the superimposed CFR of the direct link and the reflecting link. Apparently, the reflection design of ${\bm \phi}$ to achieve coherent channel combination requires the knowledge of ${\bf G}$ and ${\bf d}$. 
For practical implementation, we propose a new protocol to execute channel estimation and reflection optimization for data transmission according to the following two steps:
\begin{itemize}
\item First, based on the pilot tones of ${\bf X}$ and the pre-designed IRS reflection pattern, 
we estimate the CSI of ${\bf G}$ and ${\bf d}$; 
\item Second, based on the estimated CSI of ${\bf G}$ and ${\bf d}$, we optimize the IRS reflection ${\bm \phi}$ for data transmission.
\end{itemize}

\vspace{-0.4cm}
\subsection{Transmission Protocol}

\begin{figure}[!t]
	\centering
	\includegraphics[width=2.4in]{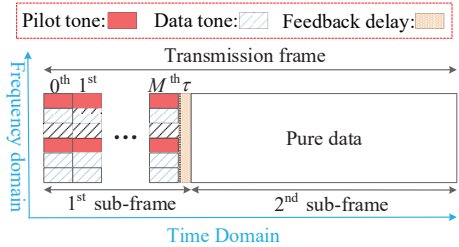}
	\caption{Illustration of the proposed transmission protocol.}
	\label{Protocol1}
	\vspace{-0.5cm}
\end{figure}

As shown in Fig. \ref{Protocol1}, one transmission frame is divided into two sub-frames in the proposed transmission protocol: the first sub-frame consists of 
$(M+1)$ consecutive pilot symbols (indexed in increasing
time order by $i\in \{0,1,\ldots, M \}$) appended with a small feedback interval $\tau$ (assumed negligible in this letter for simplicity), while the second sub-frame consists of multiple consecutive data symbols in the remaining duration of the frame. 
To resolve the superimposed CSI of ${\bf G}$ and ${\bf d}$,
each pilot symbol ${\bf X}^{(i)}$ sent by the user is associated with a pre-designed IRS reflection state
${\bm \phi}^{(i)}$, both of which are known at the AP.
Then, based on the $(M+1)$ consecutive pilot symbols and their pre-designed reflection states, i.e., $\left\{{\bf X}^{(i)}, {\bm \phi}^{(i)}\right\}_{i=0}^M$, the AP can estimate the CSI of ${\bf G}$ and ${\bf d}$, with the details given in Section~\ref{CE_int}.

With the estimated CSI of ${\bf G}$ and ${\bf d}$, our objective is to maximize the average achievable rate in the second sub-frame subject to the IRS reflection amplitude constraint, which is formulated as the following optimization problem.
\vspace{-0.2cm}
\begin{align}
\hspace{-0.2cm}\text{(P1):}~
& \underset{{\bm \phi}}{\text{max}}
& & C\left({\bm \phi}  \right)=\frac{1}{N+L_{cp}}\sum_{n =0 }^{N-1} \log_2 \left(\hspace{-0.05cm}1\hspace{-0.05cm}+\hspace{-0.05cm} \frac{P_t{\hat W}_n\left({\bm \phi}  \right) }{N\Gamma \sigma^2}\hspace{-0.05cm}\right)\hspace{-0.2cm}\label{achievable0}\\
& \text{s.t.} & &   \left|{\phi}_m\right| = 1, \quad \forall m=1, \ldots, M
\end{align}
where ${\hat W}_n\left({\bm \phi}  \right)$ is the estimated channel gain of the $n$-th sub-carrier which varies with the IRS phase-shift vector ${\bm \phi}$ and $\Gamma\ge 1$ stands for the achievable rate gap due to a practical modulation and coding scheme.
Note that to achieve the optimal solution of problem (P1), variable ${\bm \phi}$ needs to balance the channel gains $\{{\hat W}_n\left({\bm \phi}  \right)\}_{n=0}^{N-1}$ over different sub-carriers. It can be verified that problem (P1) is non-convex and thus difficult to solve optimally.
We will solve this problem in Section~\ref{IRS_Optimization} sub-optimally.
After solving problem (P1), the optimized phase-shifts are fed back to the IRS controller via a separate wireless link.
According to the feedback information, the IRS controller adjusts the phase shift of each sub-surface to achieve desired signal reflection for the data transmission in the second sub-frame. 

From the above, we see that the training overhead scales with the number of pilot symbols $(M+1)$ and/or the number of pilot tones $N_p$ in each pilot symbol and an intuitive impact of the training overhead can be envisioned as follows: with too little training the CSI is not accurate enough for reflection design and achieving high passive beamforming gain, while too much training results in less time for data transmission, both reducing the achievable rate.
Therefore, there exists a fundamental trade-off between the channel estimation overhead and reflection performance by varying the number of sub-surfaces $M$ and/or the number of pilot tones $N_p$ in each pilot symbol, as will be shown later by simulation results.
\vspace{-0.2cm}
\section{Proposed Channel Estimation and Reflection Optimization}\label{Open-Loop}
\vspace{-0.2cm}
\subsection{Channel Estimation}\label{CE_int}

As shown in Fig.~\ref{Protocol1}, a comb-type pilot scheme is applied to the first sub-frame for the purpose of channel estimation. Specifically, $N_p$ pilots are inserted in each OFDM symbol as the pilot tones indexed by
\vspace{-0.1cm}
\begin{align}\label{position}
	\mathcal{P}=\left\{0, \Delta,\ldots,(N_p-1)\Delta  \right\}
\end{align}
with $\Delta=\lfloor N/N_p\rfloor$ being the frequency spacing of adjacent pilots,
while the data tones are indexed by $\mathcal{D}=\mathcal{N} \setminus \mathcal{P}$ during the first sub-frame.
With the pilot sequence ${\bf x}_\mathcal{P}$,
the least-square estimation of CFRs on the pilot tones $\mathcal{P}$ is given by
\begin{align}\label{LS}
	{\bf r}\triangleq &\left[R_{0}, R_{1},\ldots,R_{N_p-1}\right]^T\notag\\
	=&{\bf X}_\mathcal{P}^{-1}{\bf y}_{\mathcal{P}}={\bf h}_\mathcal{P}+{\bf X}_\mathcal{P}^{-1}{\bf { v}}_\mathcal{P}
\end{align}
where ${\bf X}_{\mathcal{P}}=\text{diag} \left({\bf x}_{\mathcal{P}}\right)$ is the diagonal matrix of the pilot sequence ${\bf x}_{\mathcal{P}}$,
and ${\bf y}_{\mathcal{P}}$, ${\bf h}_\mathcal{P}$, and ${\bf { v}}_\mathcal{P}$ denote the received signals, superimposed CFRs, and equivalent AWGNs on the pilot tones ${\mathcal{P}}$, respectively.
Based on the observation of (\ref{LS}), we employ the DFT/IDFT-based interpolation to acquire/estimate the CFRs on the data tones $\mathcal{D}$.
Specifically, let ${\tilde{\bf r}} \triangleq \left[r_{0}, r_{1},\ldots,r_{N_p-1}\right]^T$ denote the $N_p$-point IDFT of ${\bf r}$ in (\ref{LS}). Accordingly, the estimate of the time-domain superimposed channel impulse response (CIR) is given by ${ \hat{\bf r}}=\sqrt{\frac{N}{N_p}} \left[{\tilde{\bf r}}\right]_{1:L}$ and
the superimposed CFR ${{\bf h}}$ is estimated by performing the $N$-point DFT on
${\hat{\bf r}}$ padded with $(N-L)$ zeros, i.e.,
\vspace{-0.2cm}
\begin{align}\label{est_h}
{{\bf {\hat h}}}=\sqrt{\frac{1}{N}}{\bf F}_{N} \left[{\hat{\bf r}}^T, {\bf 0}_{1\times (N-L)}\right]^T={{\bf h}}+{\bf {\bar v}}
\end{align}
where ${\bf F}_{N}$ denotes the $N\times N$ DFT matrix
with $\left[{\bf F}_{N}\right]_{\imath,\jmath}=e^{-j\frac{2\pi \imath \jmath }{N}}$ for $0\le \imath,\jmath\le N-1$
 and ${\bf {\bar v}}$ stands for the equivalent noise vector distributed as $ {\mathcal N_c }({\bf 0}, {\bf F}_{N} {\bf {\tilde I}} {\bf F}^H_{N} )$ with
 \vspace{-0.2cm}
\begin{align}
{\bf {\tilde I}}=\begin{bmatrix}
\frac{\sigma^2 N}{N_p P_t}{\bf I}_L & {\bf 0}_{L\times (N-L)} \\
{\bf 0}_{(N-L)\times L } & {\bf 0}_{(N-L)\times (N-L)}
\end{bmatrix}.
\end{align}

\vspace{-0.2cm}
It can be observed that based on the pilot sequence ${\bf x}_\mathcal{P}$ on the pilot tones, the superimposed CFR ${{\bf h}}$ can be estimated according to \eqref{LS} and \eqref{est_h}. To resolve the CSI of ${\bf G}$ and ${\bf d}$,
we need to design the IRS reflection pattern during the transmission of the first sub-frame.
Specifically, let ${\bm \phi}^{(i)}$ denote the IRS reflection state during the transmission of the $i$-th pilot symbol and
the corresponding estimation of the superimposed CFR ${{\bf h}}^{(i)}$ can be expressed as
\begin{align}\label{est_h2}
{{\bf {\hat h}}}^{(i)}&={\bf G} {\bm \phi}^{(i)}+ {\bf d}+{\bf {\bar v}}^{(i)}={\vec{\bf G}} {\vec {\bm \phi}}^{(i)}+{\bf {\bar v}}^{(i)}
\end{align}
where ${\vec{\bf G}}=\left[{\bf d}, {\bf G} \right]$ and $ {\vec {\bm \phi}}^{(i)}= \begin{bmatrix}1\\{\bm \phi}^{(i)}\end{bmatrix}$.
By stacking ${{\bf {\hat h}}}^{(i)}$ with $i=0,1,\ldots,M$ into
${\bf {\hat H}}=[{\bf {\hat h}}^{(0)},{\bf {\hat h}}^{(1)},\ldots,{\bf {\hat h}}^{(M)}]$ as the estimated channel matrix, we can obtain
\vspace{-0.2cm}
\begin{align}\label{est_h3}
{\bf {\hat H}}={\vec{\bf G}} { {\bm \Theta}}+{\bf {\bar V}}
\end{align}
where ${\bm \Theta}=[{\vec {\bm \phi}}^{(0)},{\vec {\bm \phi}}^{(1)},\ldots,{\vec {\bm \phi}}^{(M)}]$ denotes the IRS reflection pattern matrix by collecting all reflection states during the first sub-frame and ${\bf {\bar V}}=[{\bf {\bar v}}^{(0)},{\bf {\bar v}}^{(1)},\ldots,{\bf {\bar v}}^{(M)}]$ denotes the noise matrix.
Based on \eqref{est_h3}, the CSI of ${\bf G}$ and ${\bf d}$ is estimated as
\vspace{-0.2cm}
\begin{align}\label{est_G}
\left[{\bf {\hat d}} ~{\bf {\hat G}} \right]={\bf {\hat H}}{ {\bm \Theta}}^{-1}.
\end{align}
Generally, when the reflection pattern matrix is a full-rank matrix, i.e., ${\rm rank} \left( { {\bm \Theta}} \right)=M+1$ under the reflection amplitude constraint of $|\phi_m^{(i)}|=1$ ($\forall m=1,\ldots, M$ and $\forall i =0,1,\ldots, M$), all the required CFRs can be extracted within $(M+1)$ pilot symbols. However, the inversion operation has a complexity order of ${\cal O}((M+1)^3)$ and also may lead to considerable noise enhancement if ${ {\bm \Theta}}$ is ill-conditioned.
\rev{From \eqref{est_G}, the mean square error (MSE) of channel estimation on a sub-carrier basis is derived as
\begin{align}\label{MSE}
\varepsilon&=\frac{1}{N} \cdot {\mathbb E}\left\{  \left\|\left[{\bf {\hat d}} ~{\bf {\hat G}} \right]
-\left[{\bf { d}} ~{\bf {G}} \right]\right\|_F^{2}
\right\}
\notag \\
&=\frac{1}{ N } \cdot {\mathbb E}\left\{ 
\left\| {\bf {\bar V}} { {\bm \Theta}}^{-1}\right\|_F^{2} 
\right\}
\notag \\
&=\frac{1}{ N } \cdot {\rm tr} \left\{
\left({ {\bm \Theta}}^{-1}\right)^H {\mathbb E} \left\{{\bf {\bar V}}^H {\bf {\bar V}} \right\}{ {\bm \Theta}}^{-1}  \right\}
\notag \\
&\stackrel{(a)}{=}\frac{\sigma^2N L}{ N_p P_t}\cdot {\rm tr} \left\{ \left({\bm \Theta}^H{\bm \Theta} \right)^{-1} \right\}.
\end{align}
where the equality of $(a)$ holds as ${\mathbb E} \left\{{\bf {\bar V}}^H {\bf {\bar V}} \right\}=\frac{\sigma^2N^2 L}{N_p P_t}{\bf I}_{M+1}$. To minimize the variance of the channel estimation error, the matrix ${\bm \Theta} $ is required to satisfy ${\bm \Theta}^H{\bm \Theta}=(M+1) {\bf I}_{M+1}$, which implies that the reflection pattern is the orthogonal matrix with each entry satisfying the unit-modulus constraint during the first sub-frame. In particular, the reflection pattern using the $(M+1)\times (M+1)$ DFT matrix ${\bf F}_{M+1}$ can meet this requirement and achieve the minimum MSE in \eqref{MSE} as $\varepsilon_{\rm min}=\frac{\sigma^2N L}{N_p P_t}$, where $\left[{\bf F}_{M+1}\right]_{\imath,\jmath}=e^{-j\frac{2\pi \imath \jmath }{M+1}}$ with $0\le \imath,\jmath\le M$. Moreover, it is worth pointing out that ${\bf F}_{M+1}^{-1}=\frac{1}{M+1}{\bf F}_{M+1}^H$, which avoids the inversion operation for achieving lower complexity.}


From the above, we see that the pre-designed reflection pattern ${\bm \Theta}$ can be regarded as a new pilot pattern, together with the user pilot sequence ${\bf X}_\mathcal{P}$ to achieve the channel estimation of ${\bf G}$ and ${\bf d}$.
\vspace{-0.4cm}
\subsection{Reflection Optimization}\label{IRS_Optimization}


Based on the estimated CSI of ${\bf G}$ and ${\bf d}$ in Section \ref{CE_int}, 
the channel gain of each sub-carrier is given by
\vspace{-0.2cm}
\begin{align}\label{MRC}
{\hat W}_n\left({\bm \phi}  \right)= \left|\sum_{m=1}^{M} { \phi}_m {\hat G}_{m,n}
+ {\hat D}_{n} \right|^2, ~~ n=0,\ldots,N-1
\end{align}
each of which depends on the phase-shift vector ${\bm \phi}$. 
We aim to optimize the IRS reflection ${\bm \phi}$ for maximizing the average achievable rate in \eqref{achievable0}, which, however, is non-concave over ${\bm \phi}$ and thus difficult to maximize optimally.
Alternatively, we consider to maximize the rate upper bound of \eqref{achievable0}, which is given by (based on the Jensen's inequality) 
\vspace{-0.2cm}
\begin{align}\label{upper_bound}
\hspace{-0.1cm}C\left({\bm \phi}  \right) \le\hspace{-0.1cm}
\frac{N}{N+L_{cp}} 
\log_2 \hspace{-0.1cm}\left(\hspace{-0.1cm}1+ \frac{1}{N}\sum_{n =0 }^{N-1}\frac{P_t{\hat W}_n\left({\bm \phi}  \right) }{N\Gamma \sigma^2} \right)
\end{align}
and formulate the following optimization problem (with constant/irrelevant terms omitted for brevity).
\vspace{-0.2cm}
\begin{align}
\text{(P2):}~
& \underset{{\bm \phi}}{\text{max}}
& & \sum_{n=0}^{N-1}\left|\sum_{m=1}^{M} { \phi}_m {\hat G}_{m,n}
+ {\hat D}_{n} \right|^2 \label{OP2obj}\\
& \text{s.t.} & &   \left|{\phi}_m\right| = 1, \quad \forall m=1, \ldots, M \label{OP2con1}
\end{align}
which turns out to be the maximization of the sum channel power gain at the receiver.
Similar to \cite{yang2019intelligent}, 
SDR method can be applied to solve problem (P2) sub-optimally.
Although the SDR method achieves close-to-optimal performance in \cite{yang2019intelligent}, its complexity for solving (P2) can be shown in the order of ${\cal O}((M+1)^6)$, which is practically costly for large values of $M$. \rev{Hence, we propose in this letter a low-complexity alternative method to solve problem (P2) sub-optimally by exploiting the time domain property.
Specifically, the objective function of (\ref{OP2obj}) can be transformed into the time domain as (based on the Parseval's theorem)}
\vspace{-0.2cm}
\begin{align}
\sum_{l=0}^{L-1}  \left|\sum_{m=1}^{M} { \phi}_m {\hat g}_{m,l}
+ {\hat d}_{l} \right|^2
\end{align}
where ${\hat g}_{m,l}$ denotes the $l$-th tap of the estimated CIR for the cascaded reflecting link associated with the $m$-th IRS sub-surface and ${\hat d}_{l}$ denotes the $l$-th tap of the estimated CIR for the direct link. \rev{Note that in typical wireless environment, we have $L\leq L_{cp}\ll N$, which implies that the channel power is much more concentrated in the time domain than that in the frequency domain. Motivated by this, we propose to find the strongest CIR gain with respect to the tap index $l$, i.e,} 
\vspace{-0.2cm}
\begin{align}\label{largest_CIR}
\breve{l}=\arg \max_{l\in \{0,\ldots, L-1\}  }   \left|\sum_{m=1}^{M}   \left| {\hat g}_{m,l}\right|
+ \left|{\hat d}_{l}\right| \right|^2
\end{align}
\rev{and align the reflection phase shifts to the strongest CIR as}
\begin{align}\label{phi_m}
{\breve \varphi}_m=-\angle {\hat g}_{m,\breve{l}}+ \angle{\hat d}_{\breve{l}},\quad\quad  m=1,\ldots, M
\end{align} 
which is referred to as the strongest-CIR maximization (SCM) method.
\rev{It is worth pointing out that maximizing the strongest time-domain CIR with phase alignment in \eqref{phi_m} is practically effective since its power is equally spread out in the frequency domain, which is beneficial to all sub-carriers according to the Parseval's theorem.}
	In particular, when $L=1$, the phase shifts given in (\ref{phi_m}) are optimal to both the problems (P1) and (P2). 

\section{Numerical Results and Discussions}
In this section, we provide simulation results to demonstrate the effectiveness of our proposed channel estimation and reflection optimization methods.
We consider a uniform square array for the IRS, which consists of $K=12\times 12=144$ reflecting elements with half-wavelength spacing.
The path loss exponents of the user$\rightarrow$AP, user$\rightarrow$IRS, IRS$\rightarrow$AP links
are set as $3.5$, $2.4$, and $2.2$, respectively, and
the path loss at the reference distance of $1$ meter (m) is set as $30$ dB for each individual link.
The distance between the IRS and AP is $50$ m and the user lies on a horizontal line at a distance of $2$ m in parallel to that connecting the IRS and the AP, similarly as in \cite{Wu2019TWC}.
The transmission frame consists of $150$ OFDM symbols, 
where each OFDM symbol consists of $N =64$ sub-carriers and is appended by a CP of length $L_{cp}=8$. 
The Zadoff-Chu sequence \cite{Polyphase1972Polyphase}
is employed as the pilot sequence during the first sub-frame.
The frequency-selective Rician fading channels with delay spread
of $L=6$ taps are considered for both direct link and reflecting link,
where the first tap is set as the deterministic line-of-sight (LoS) component and the remaining taps are non-LoS components following the Rayleigh fading distribution, with $\eta$ being the ratio of the total power of non-LoS components to that of LoS component.
Other parameters are $\Gamma=9$ dB, $\sigma^2=-80$ dBm,
 and the number of randomizations in the SDR method \cite{yang2019intelligent} for solving (P2), which is set as $100$.
The ON/OFF-based channel estimation method adopted in \cite{yang2019intelligent} is considered for comparison, where the direct channel is estimated with all sub-surfaces turned OFF
and the reflecting link is estimated with one out of $M$ sub-surfaces turned ON sequentially.



\begin{figure}[!t]
	\centering
	\includegraphics[width=2.7in]{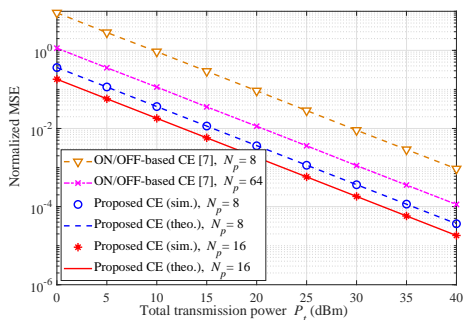}
	\caption{\rev{Normalized MSE versus transmit power $P_t$.}}
	\label{CE_MSE_comp}
	\vspace{-0.3cm}
\end{figure}
In Fig. \ref{CE_MSE_comp}, we examine the channel estimation performance in terms of MSE $\varepsilon$ normalized to the channel gain, with $M=12$, $\eta=0.5$, and the user-AP horizontal distance of $45$ m. 
It can be observed that the theoretical analysis of MSE in \eqref{MSE} is in perfect agreement with the simulation results for our proposed channel estimation method.
By doubling the number of pilot tones $N_p$, our proposed channel estimation method achieves about $3$ dB power gain, which also corroborates the accuracy of \eqref{MSE} as $10\log_{10}2\approx 3$ dB.
Moreover, under the same number of pilot tones $N_p=8$, our proposed channel estimation method achieves up to $14$ dB
gain over the ON/OFF-based counterpart in \cite{yang2019intelligent}.
Such a large performance gap is attributed to the reflection power loss and noise enhancement in the ON/OFF-based channel estimation method.
Therefore, the choice of IRS reflection pattern has a significant impact on the MSE performance of channel estimation.

\begin{figure}[!t]
	\centering
	\includegraphics[width=2.7in]{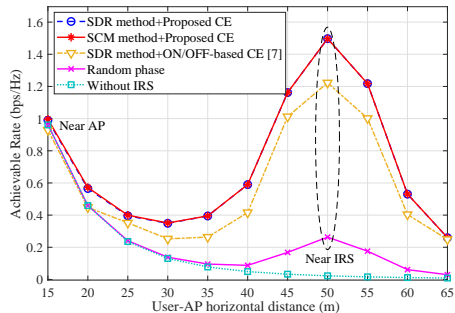}
	\caption{\rev{Achievable rate versus user-AP horizontal distance.}}
	\label{rate_dist}
	\vspace{-0.3cm}
\end{figure}
In Fig. \ref{rate_dist}, we compare the achievable rates of different schemes versus the user-AP horizontal distance, with $M=12$, $N_p=64$, $\eta=0.5$, and $P_t=0$ dBm.
It is observed that those schemes aided with IRS outperform the one without IRS, especially when
the user locates in the vicinity of the IRS. 
Compared to the SDR method,
our proposed SCM method achieves nearly the same performance but with a much lower complexity by exploiting the strongest channel path resolved in the time domain. On the other hand, we see that using the same SDR method for reflection optimization, our proposed channel estimation method yields a significant gain over the ON/OFF-based counterpart.

\begin{figure}[!t]
	\centering
	\includegraphics[width=2.7in]{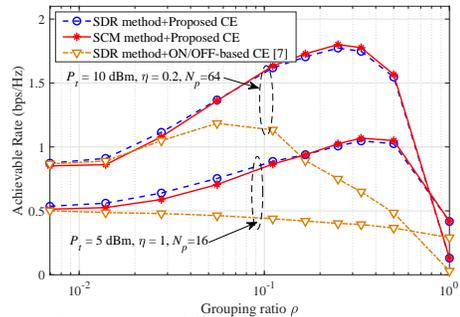}
	\caption{\rev{Achievable rate versus IRS grouping ratio $\rho$.}}
	\label{rate_grouping}
	\vspace{-0.3cm}
\end{figure}
\rev{Fig.~\ref{rate_grouping} shows the achievable rates of different schemes versus the IRS grouping ratio defined by $\rho \triangleq M/K$, with the user-AP horizontal distance of $45$ m.
The pilot overhead ratio (in terms of the number of pilot symbols $(M+1)$ and the number of pilot tones $N_p$) for channel estimation is taken into account for plotting the achievable rate performance.} It is observed that using the proposed channel estimation, both reflection optimization methods achieve comparable performance, while the SDR and SCM methods show slightly superior performance to each other at low and high grouping ratios, respectively. 
Such phenomenon can be explained by the fact that when channel power is dominated by the LoS component, maximizing the strongest time-domain CIR gain with phase alignment in \eqref{phi_m} is more effective at high grouping ratio.
\rev{Moreover, owing to the improved channel estimation accuracy and the higher aperture gain during the first sub-frame, our proposed schemes achieve much better rate performance trade-off between the channel estimation overhead and IRS reflection performance by varying $M$ (or equivalently $\rho$) and/or $N_p$, as compared to the ON/OFF-based channel estimation in \cite{yang2019intelligent} even with the SDR-based reflection optimization for data transmission.}
\section{Conclusions}
In this letter, we have proposed a practical transmission protocol to execute channel estimation and reflection optimization for the IRS-enhanced OFDM system.
Under the unit-modulus constraint, we have designed a novel reflection pattern for channel estimation and optimized the reflection coefficients with a low-complexity SCM method.
Simulation results have verified the superior performance of our proposed methods over the existing schemes.  

%
%
%
%
%
%
%
%

\ifCLASSOPTIONcaptionsoff
  \newpage
\fi

\bibliographystyle{IEEEtran}
\bibliography{IRS_OFDM}

\begin{thebibliography}{1}
\providecommand{\url}[1]{#1}
\csname url@samestyle\endcsname
\providecommand{\newblock}{\relax}
\providecommand{\bibinfo}[2]{#2}
\providecommand{\BIBentrySTDinterwordspacing}{\spaceskip=0pt\relax}
\providecommand{\BIBentryALTinterwordstretchfactor}{4}
\providecommand{\BIBentryALTinterwordspacing}{\spaceskip=\fontdimen2\font plus
\BIBentryALTinterwordstretchfactor\fontdimen3\font minus
  \fontdimen4\font\relax}
\providecommand{\BIBforeignlanguage}[2]{{%
\expandafter\ifx\csname l@#1\endcsname\relax
\typeout{** WARNING: IEEEtran.bst: No hyphenation pattern has been}%
\typeout{** loaded for the language `#1'. Using the pattern for}%
\typeout{** the default language instead.}%
\else
\language=\csname l@#1\endcsname
\fi
#2}}
\providecommand{\BIBdecl}{\relax}
\BIBdecl

\bibitem{qingqing2019towards}
Q.~Wu and R.~Zhang, ``Towards smart and reconfigurable environment: Intelligent
  reflecting surface aided wireless network,'' \emph{IEEE Commun. Mag.}, doi:
  10.1109/MCOM.001.1900107, Nov. 2019.

\bibitem{Renzo2019Smart}
M.~Di~Renzo \emph{et~al.}, ``Smart radio environments empowered by
  reconfigurable {AI} meta-surfaces: An idea whose time has come,''
  \emph{EURASIP J. Wireless Commun. Netw.}, vol. 2019:129, May 2019.

\bibitem{Huang2018Achievable}
C.~{Huang}, A.~{Zappone}, M.~{Debbah}, and C.~{Yuen}, ``Achievable rate
  maximization by passive intelligent mirrors,'' in \emph{Proc. IEEE ICASSP},
  Apr. 2018, pp. 3714--3718.

\bibitem{Wu2019TWC}
Q.~Wu and R.~Zhang, ``Intelligent reflecting surface enhanced wireless network
  via joint active and passive beamforming,'' \emph{IEEE Trans. Wireless
  Commun.}, vol.~18, no.~11, pp. 5394--5409, Nov. 2019.

\bibitem{Mishra2019Channel}
D.~{Mishra} and H.~{Johansson}, ``Channel estimation and low-complexity
  beamforming design for passive intelligent surface assisted {MISO} wireless
  energy transfer,'' in \emph{Proc. IEEE ICASSP}, May 2019, pp. 4659--4663.

\bibitem{he2019cascaded}
Z.-Q. He and X.~Yuan, ``Cascaded channel estimation for large intelligent
  metasurface assisted massive {MIMO},'' \emph{IEEE Wireless Commun. Lett.},
  doi: 10.1109/LWC.2019.2948632, Oct. 2019.

\bibitem{yang2019intelligent}
Y.~Yang, B.~Zheng, S.~Zhang, and R.~Zhang, ``Intelligent reflecting surface
  meets {OFDM}: Protocol design and rate maximization,'' \emph{arXiv preprint
  arXiv:1906.09956}, 2019.

\bibitem{Polyphase1972Polyphase}
D.~{Chu}, ``Polyphase codes with good periodic correlation properties,''
  \emph{IEEE Trans. Inf. Theory}, vol.~18, no.~4, pp. 531--532, Jul. 1972.

\end{thebibliography}

\end{document}